\documentstyle[preprint,aps,epsf]{revtex}
\tightenlines

\def\MeV{\nobreak\,\mbox{MeV}}
\def\GeV{\nobreak\,\mbox{GeV}}

\def\epm{e^+e^-}
\newcommand{\be}{\begin{eqnarray}}
\newcommand{\ee}{\end{eqnarray}}
\newcommand{\bes}{\begin{eqnarray*}}
\newcommand{\ees}{\end{eqnarray*}}

\newcounter{dafigcounter}
\newcommand{\pfig}[3]{
 \refstepcounter{dafigcounter}
 \begin{minipage}[t]{#2}
  \begin{center}
   {\epsfxsize=#2 \mbox{\epsffile{#1.eps}}}
  \end{center}
  \label{#1}
  \small \bf Fig.~\thedafigcounter\rm\ #3
 \end{minipage}
}

\begin{document}
\draft
\title{Transport calculation of dilepton production at
ultrarelativistic energies}
\author{C.~Ernst$^{a}$, S.~A.~Bass$^{b}$, S.~Soff$^a$,
H.~St\"ocker$^a$ and W.~Greiner$^a$
\\
}
\address{
$^a$Institute for Theoretical Physics, University of
Frankfurt,
Robert-Mayer-Strasse 8-10, D-60054 Frankfurt, Germany \\
$^b$Department of Physics, Duke University, Durham,
North Caroline 27708-0305, USA
}

%
\maketitle
%

\begin{abstract}
Dilepton spectra are calculated within the microscopic
transport model UrQMD and compared to data from the CERES
experiment. The invariant mass spectra in the
mass region between 300 and 600~MeV depend strongly on the
mass dependence of the $\rho$ meson decay width which is
not sufficiently determined by the Vector Meson Dominance
(VMD) model. A consistent explanation of both the recent Pb+Au
data and the proton induced data can be given
without additional medium effects.
\end{abstract}
\vspace{1cm}

\section{Introduction}

Recently, electromagnetic radiation in form of lepton pairs
has been observed at CERN in the CERES experiment. A strong
enhancement above the cocktail of hadronic decays has been
reported. This is interesting, because dileptons may escape
nearly undisturbed from the hot and dense zone which is formed in
the heavy ion collision. Dileptons could probe the
intermediate stage of these interactions.  On the other
hand, hadrons are thought to probe the later freeze-out
stages. This folklore may not be entirely correct: A number
of investigations indicate that hadronic signals such as
strangeness and entropy, as well as flow are developped in the
early phase, while a large fraction of the observed
dileptons is simply due to post-freeze-out decays
("feeding") of isolated hadrons.

The fact that also dileptons are emitted from various
sources and during the entire reaction makes the
interpretation of this signal not as easy as
hoped for. Electromagnetic probes are definitely not a pure
signal from the most interesting high density stage only!
Rather, their spectrum is composed of different
contributions integrated over the full interaction
dynamics. Transport models need to be applied to decompose
the dilepton cocktail into its various individual
contributions.

The most discussed feature of the data is the enhancement
of intermediate mass lepton pairs $(300\MeV < M_{ee} <
700\MeV)$ as compared to the hadronic cocktails based on
primary collisions.  While the enhancement at BEVALAC
energies is still not understood
\cite{porter97a,bratko97a,bratko98a,ernst98a}, first predictions for Pb+Au
collisions from transport models including secondary
production of dileptons (e.g.
$\pi\pi\to\epm$) indicated only a small (if any) deviation
from the CERES data \cite{winckelmann96a,koch96a}.
A large variety of attempts has been made
to explain this deviation. The two most popular of
them are the hypothesis of
lowering vector meson masses (the Brown-Rho scaling)
\cite{brown91a,hatsuda92a,ligq96a} and the collisional
broadening of the spectral functions in the hadronic
environment \cite{rapp97a,cassing97a}. 

Here we analyse the recent CERES data on the multiplicity
and $p_t$ of the lepton pairs. 
First, we give a survey of the dilepton production in the
framework of the ultrarelativistic quantum molecular
dynamics model, UrQMD \cite{bass98a}. The 
present paper demonstrates that a 
state-of-the-art transport model based on hadronic and
string degrees of freedom can predict the present data
without assuming phase transitions or unconventional in-medium effects. 
This compares well to recent results from a simple transport
model \cite{koch99a}.

\section{Description of the model}

Dilepton production in the framework of the UrQMD model has
been considered in some detail in ref. \cite{ernst98a}. In
this model dileptons are produced perturbatively, mainly in
hadronic decays. Dalitz decays of neutral mesons and direct
decays of vector mesons are taken into account. The
incoherently summed $pn$ bremsstrahlung as well as the
$\Delta$(1232) Dalitz decay are of minor importance at
CERES energies because the system is dominated by mesonic
degrees of freedom. All dilepton decay widths have been
specified in \cite{ernst98a}.

For the vector meson decay width $\Gamma_{\rho\to\epm}(M)$
we use the typical $M^{-3}$ dependence which results from
vector meson dominance (VMD model). There seems to be some
inconsistency between the original formulation of VMD (see
e.g. Sakurai \cite{sakurai}), which results in a $M^{-3}$
dependence for $\Gamma_{\rho\to\epm}$ and the
extended VMD from Kroll, Lee and Zumino \cite{kroll67a}
implying a dilepton decay width proportional to $M$
(see e.g. ref. \cite{klingl96a}). The
discrepancy between these two approaches seems to be as
$M^{4}$! However, if one considers the channel
$\pi\pi\to\epm$, the extended VMD requires also a direct
coupling between the electromagnetic and the hadronic channel.  If
this term is included (as well as the interference), one
ends up with the overall $M^{-3}$ dependence again.

In the extended VMD the direct term can be
motivated only by a $\pi\pi\gamma$ coupling. Therefore it is
not clear
how to treat the dilepton decay of $\rho$ mesons which
themselves come
from heavy resonance decays or even string fragmentation.
Similar problems arise for the $\omega$ and $\phi$ decays.
However, due to their small total widths, this effect
does not manifest itself in the dilepton mass spectra and is thus
not discussed separately here. In the UrQMD model the
following procedure is applied:

For $\rho$ mesons which are produced by
$\pi\pi$-annihilations or in decays of baryonic or mesonic
resonances the dilepton decay width scales like $M^{-3}$. The 
dilepton width for $\rho$ mesons from string fragmentation 
is chosen according to the direct
term of the extended VMD model which scales like $M$. 
Other approaches are possible and yield different results:
see discussion below.

\section{Results}

The undisturbed dilepton mass spectrum for
proton on beryllium collisions at 450~GeV incident energy
is shown in fig.~\ref{pbeno}. This 'naked' spectrum
could be verified only with an ideal detector
covering the $4\pi$ phase space angle and having a perfect
momentum resolution (the bin width is 8~MeV).  It shows all
details of the model
calculation. The CERES data \cite{agakishiev97a} are drawn
for orientation only and should not be directly compared to the
actual calculation. One sees, however, that roughly two
orders of magnitude are lost due to the acceptance.

At low invariant masses the overwhelming background is due
to the Dalitz decays of the long-living mesons $\pi^0$ and
$\eta$. Between 500 and 750~MeV the $\omega$ Dalitz and the
direct $\rho$ meson decays dominate the spectrum. At higher
energies the characteristic peaks of the direct $\omega$
and $\phi$ decays overlap the $\rho$ decay. The $\eta'$
Dalitz decay or baryonic channels like $pn$ bremsstrahlung
play only a subordinate role. The asymmetrical shape of the
$\rho$ mass distribution is due to phase space limitations
and especially due to the
$M^{-3}$ scaling of the decay width. At invariant masses around 1.2~GeV
there is a drop in the $\rho$ distribution. This is
due to a limitation of the model not to create resonances
from string fragmentation with masses exceeding the pole
mass more than three widths ($M < M_R + 3\Gamma_R$). 

For comparison with the experimental data from the CERES
collaboration, one needs to correct the model calculations
for the limited detector acceptance and momentum
resolution. Because the detector covers only a small
geometric region, every detected particle must have a
certain angle to the beam axis. This requires the pseudo
rapidity of the lepton $\eta_{e^\pm}$ to be $2.1 \le
\eta_{e^\pm} \le 2.65$. A cut on small opening angles
$\Theta_{\epm} > 0.035~\mbox{mrad}$ of the
electron-positron pair has to be matched because close
pairs cannot be identified by the RICH detectors. The
individual leptons which have passed the initial RICH
detectors are then deflected in a magnetic field with a
momentum resolution parameterised as
\be
\frac{\Delta p}{p} = \sqrt{\alpha^2 + (\beta p)^2}\; ,
\ee
where $\alpha$ and $\beta$ depend on the detector setup and
are listed in table 1 \cite{lenkeit}. The
momenta of the electron and positron are washed out
accordingly with a Gaussian distribution function of width
$\frac{\Delta p}{p}$. Finally, these tracks have to
survive a transverse momentum cut, which is typically $p_t > 50\MeV$
for the proton induced and $p_t > 200\MeV$ for the ion
induced reactions.

In fig.~\ref{pbe} we present the dilepton mass spectra for
proton on beryllium collisions including the CERES
acceptance and resolution. As compared to fig.~\ref{pbeno}
the momentum resolution has washed out the peak structures
and the low mass pairs have been suppressed due to the
acceptance. As an effect of the finite resolution one can
see also contributions from the direct $\rho^0$ decay below
the two pion threshold. The transport calculation slightly
overestimates the data around the vector meson poles but
seems to be consistent with the overall data set. At
invariant masses higher than 1.3~GeV the data possibly indicate a
need for additional sources like direct production in
meson+meson collisions (see e.g.
\cite{li98a}) which were not considered in the present
calculation.

The UrQMD result for the heaviest system measured by CERES
can be found in fig.~\ref{pbau} together with the data from
the '96 run \cite{lenkeit}. These default calculations do
not show the typical trend to underestimate the data in the
region around 400~MeV. Like in a number of other free or
conventional transport calculations the data point at
780~MeV, just at the $\rho/\omega$ mass, is slightly
overestimated \cite{rapp97a,cassing97a,koch99a}.
This has been interpreted that 'some of the signal should
be distributed away' by medium effects. However, in view of
the good agreement of the conventional calculation at lower
masses one must ask, where this strength should go.
Furthermore, we have observed that introducing collision
widths results in a significant enhancement of vector meson
production, which is not compatible to the data.

Two data points are missed at 140~MeV and 180~MeV. By
lowering the $p_t$ cut to less than 200~MeV this dip can be
artificially removed on cost of overshooting the data points at 63~MeV
and 88~MeV. Our interpretation of this observation is that
the UrQMD $p_t$-distribution of pions and etas is too soft
to be compatible to this data.  Furthermore the spectral
density of the $\rho$ meson
in nuclear matter might extend to below $2 m_\pi$
because the $\rho$ may decay into a nucleon nucleon-hole pair. This
contribution is neglected in all on-shell
transport models like UrQMD, but might fill in this gap.

The results of the transport simulation depend on the
details of the VMD model (fig.~\ref{pbauvdm}). In the
default calculation, the dilepton decay width of $\rho$
mesons
scales like $M^{-3}$ if they are from hadronic
sources and like $M$ for
$\rho$'s from strings (see the
discussion above). If no VMD was considered at all, i.e.
assuming constant dilepton decay widths for the $\rho$, a
clear enhancement of the data over the model calculation
can be observed (dashed line in fig.~\ref{pbauvdm}).
However, as pointed out above, neglecting the
$M^{-3}$-dependence of the
$\rho$ meson dilepton decay width is not a valid
assumption. In the hydrodynamical calculations of ref.
\cite{sollfrank97a} the $M^{-3}$ factor is missed, which
might be the reason for their underestimation of 
the CERES data.

On the other hand, one can assume 'total VDM' by requiring
that the vector mesons which were produced in string
fragmentation scale like the 'hadronic' $\rho$'s with
$M^{-3}$. This leads to an overestimation of the Pb+Au data
(see dotted line in fig.~\ref{pbauvdm}) and also the p+Be
data are overestimated. From a theoretical point of view
this approach has to be considered as an upper limit to the
contribution of dilepton emission from $\rho$ mesons from
string fragmentation. Contrary, the default calculation
provides a lower limit for a realistic dilepton spectrum.
The range between both approaches is the result of our
uncertainty how to treat dilepton production from string
fragmentation.

The recent lead data have also been analysed for the
transverse momenta of the lepton pairs (i.e. the virtual
photons). The idea was to cut on pairs emitted from
low-$p_t$ vector mesons for which the interaction with the
surrounding medium should be strongest. Consequently,
in-medium effects would result in a stronger enhancement
of the low-$p_t$ data set over conventional models. The
results of the corresponding UrQMD analysis are shown in
fig.~\ref{pbaupt} in comparison to the CERES data
\cite{lenkeit}. 

Note, first of all that the dip below
$M_{\epm} = 200\MeV$ which was already observed in
fig.~\ref{pbau} can be found only in the high
$p_t$ region (lower frame). This indicates once more that
UrQMD predicts less high-$p_t$ $\pi^0$ and $\eta$ mesons.

In the $\rho$ region a reasonable agreement between
simulation and data can be found both for the low-$p_t$
(upper frame) and the high-$p_t$ sample. In particular the
low-$p_t$ data are consistent with the present UrQMD
calculation -- without a need for additional medium
effects.


\section{Conclusion}

In summary, the p+Be and the Pb+Au CERES data can be
consistently explained by the UrQMD transport simulations.
A reasonably good agreement was found especially in the
region of $0.3\GeV < M < 0.6\GeV$. It was discussed that
the parameterisation of the vector meson decay width is not
constrained from first principles. The variation between
different approaches for
$\Gamma_{\rho\to\epm}(M)$ manifests itself in dilepton
spectra clearly underestimating 
or slightly overshooting the CERES data. Further efforts in
theory are required to provide a microscopic scheme
for $\Gamma_{\rho\to\epm}(M)$. Measurements with higher
precision could help to identify the $\omega$ peak in the
invariant mass spectra and thus pin down the shape of the
$\rho$ meson. In addition more experimental information on the
elementary channels p+p, $\pi$+p, especially a separate
measurement of the Dalitz contribution (e. g. with TAPS),
could be used to verify the different realisations of the VMD.

\section*{Acknowledgments}
The authors want to thank W.~Cassing, B.~Friman and J.~Knoll
for helpful discussions.
This work was supported in parts by
Graduiertenkolleg Theoretische und Experimentelle
Schwerionenphysik, GSI, BMBF, DFG, J.~Buchmann Foundation,
A.~v.~Humboldt Foundation and D.~O.~E. grant
DE-FG02-96ER40945.


\begin{table}
\begin{center}
\begin{tabular}{|r|c|c|}
\hline
& $\alpha$ & $\beta ~ (\MeV)^{-1}$ \\
\hline
p + Be (1994) & 0.043 & 0.053\\
Pb + Au (1996) & 0.035 & 0.023\\
\hline
\end{tabular}
\end{center}
\caption{Parameters of the CERES momentum
resolution (see text).}
\end{table}

\begin{figure}[thb]
\begin{center}
\pfig{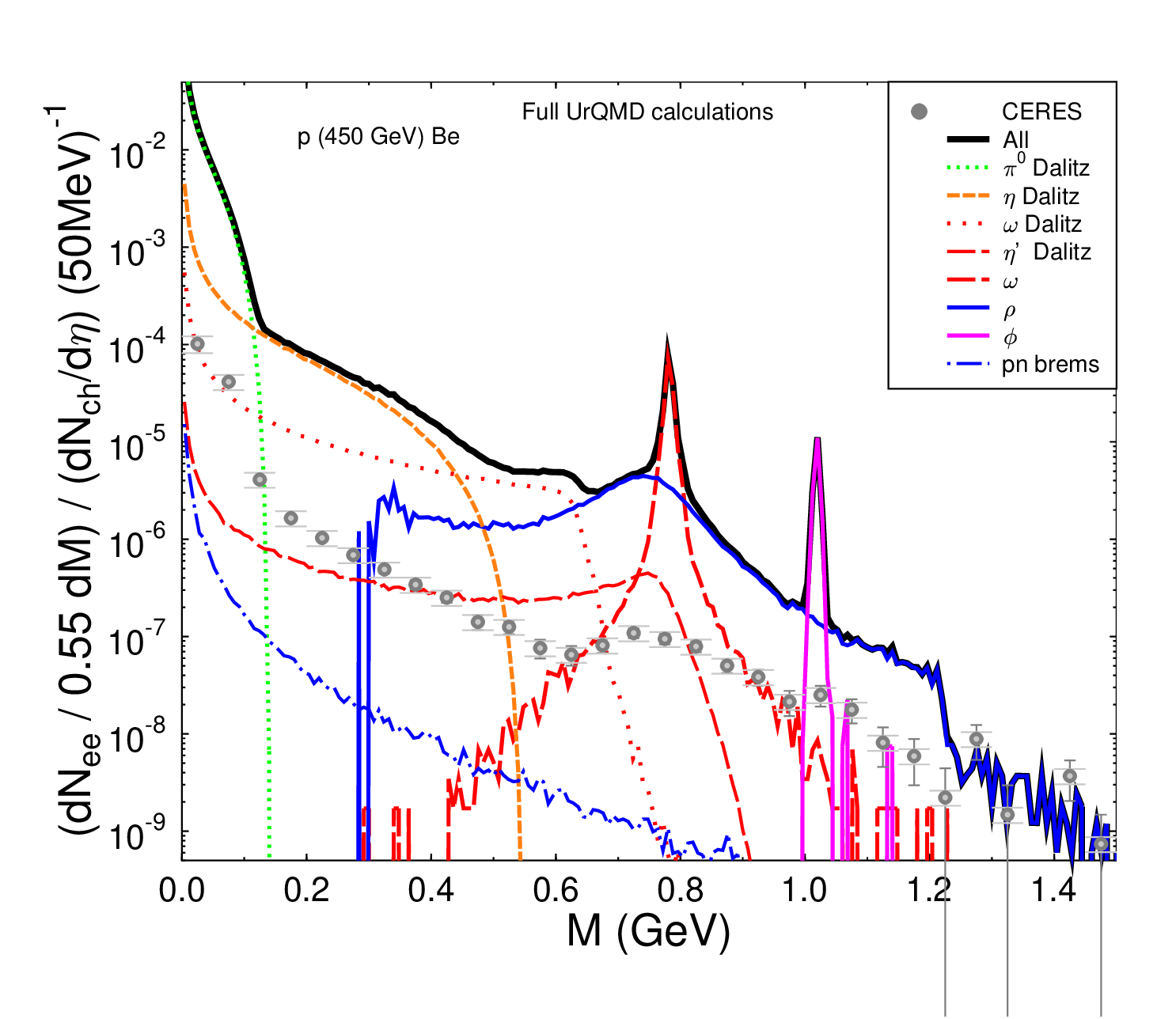}{12cm}
{'Ideal' dilepton spectrum of p + Be collisions at 450~GeV.
The UrQMD results are not filtered and
the data points are for orientation only.}
\end{center}
\end{figure}
\begin{figure}[thb]
\begin{center}
\pfig{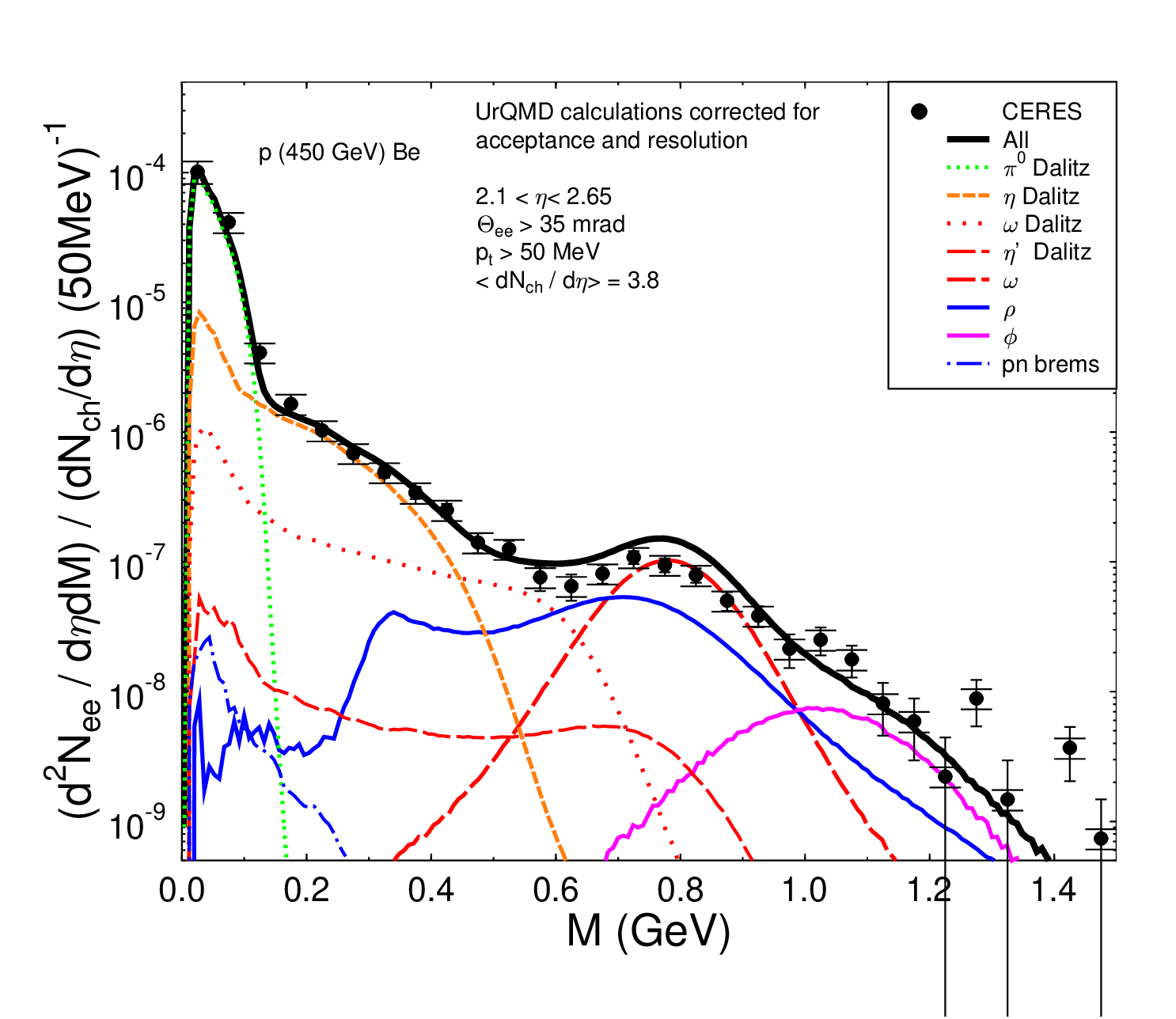}{12cm}
{Dilepton spectrum of p + Be collisions at 450~GeV.
UrQMD calculation, corrected for acceptance and resolution,
in comparison to data \cite{agakishiev97a}}.
\end{center}
\end{figure}
\begin{figure}[thb]
\begin{center}
\pfig{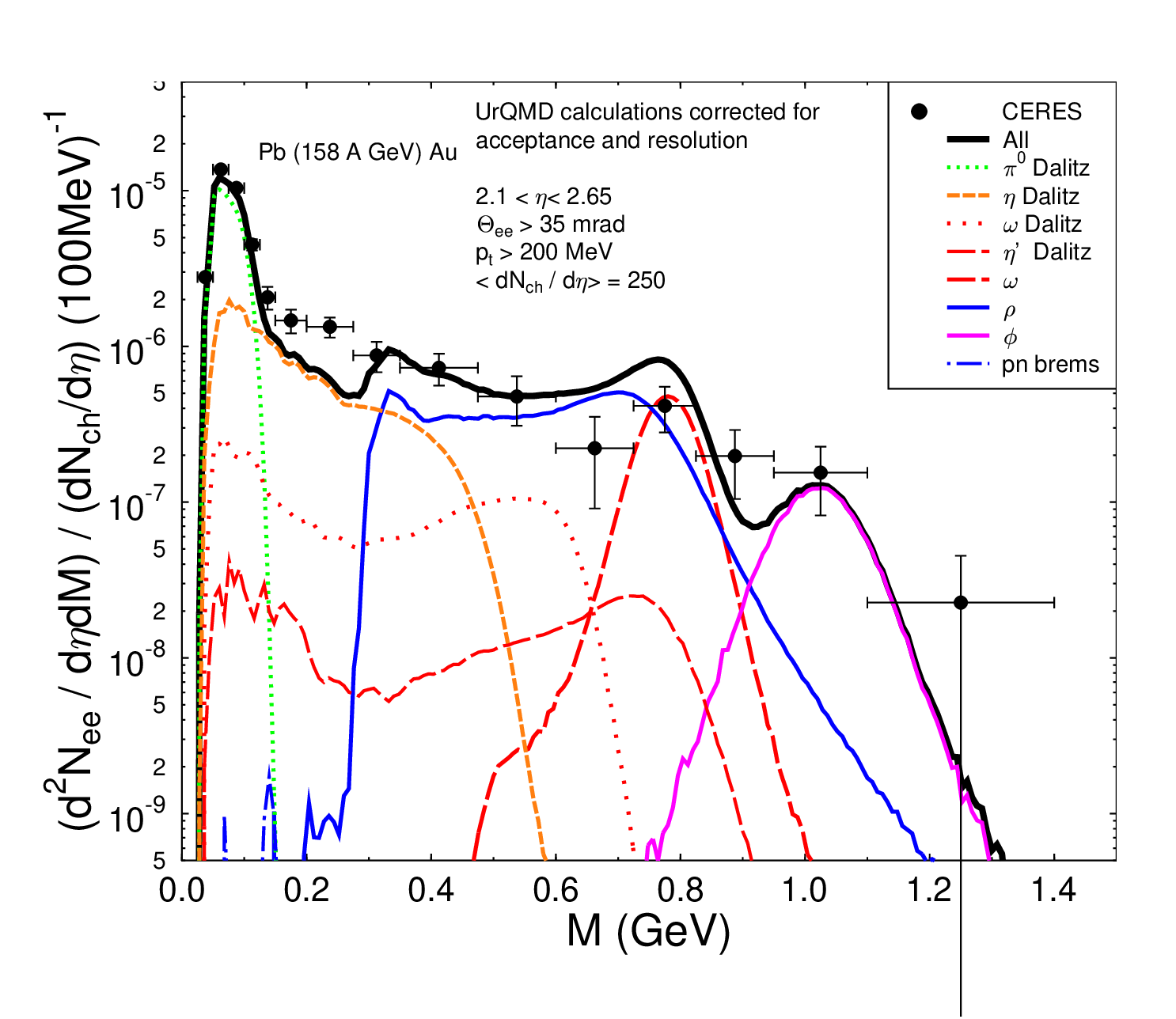}{14cm}
{Dilepton spectrum of Pb + Au collisions at 158~A~GeV.
The '96 CERES data \cite{lenkeit} compared to filtered
UrQMD calculations.}
\end{center}
\end{figure}
\begin{figure}[thb]
\begin{center}
\pfig{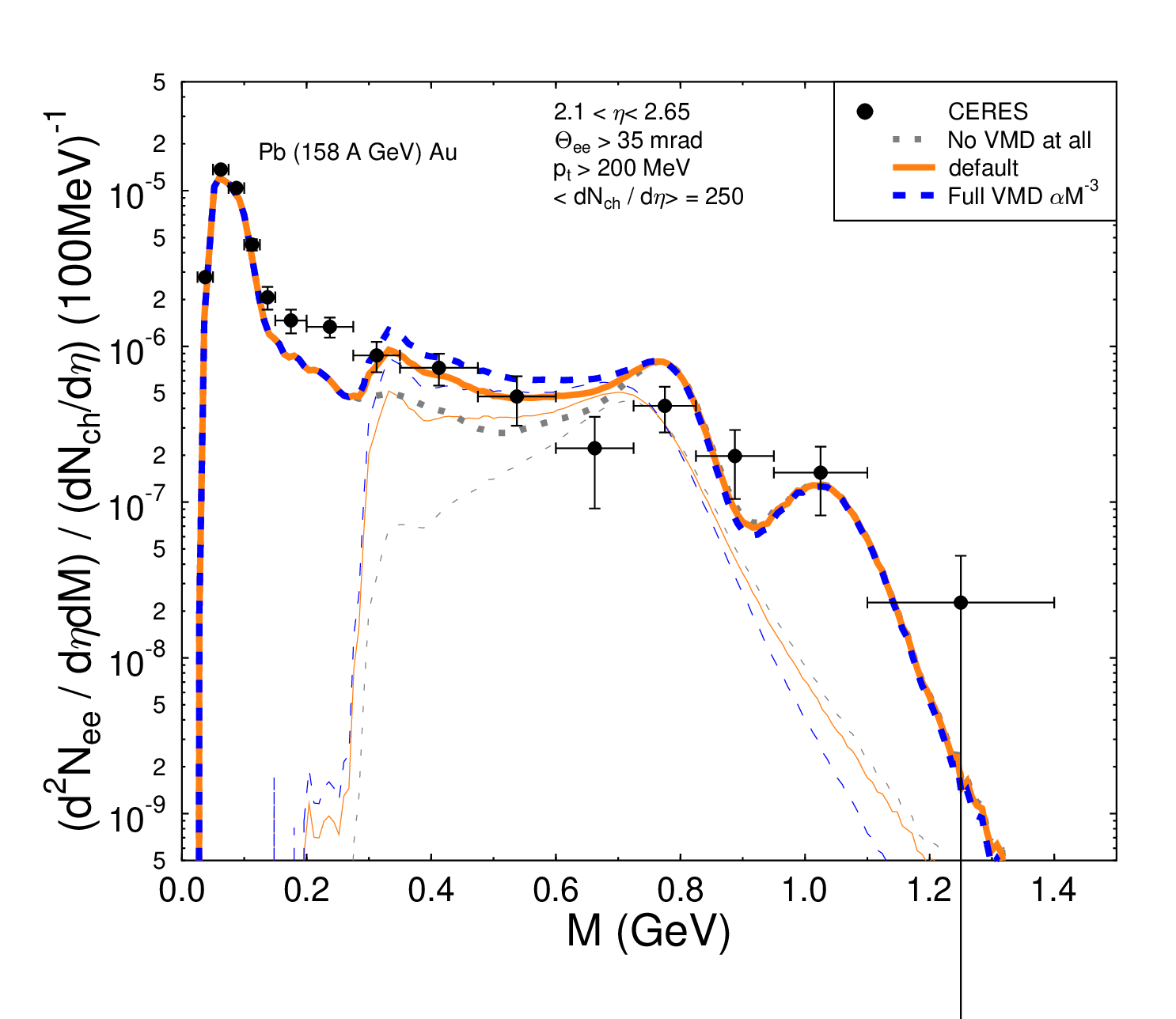}{14cm}
{Effect of the mass dependence of the $\rho$ meson dilepton
decay width. Thin lines represent the mass distributions in
the $\rho$ channel while fat lines are the total
spectra. (See text for more details.)}
\end{center}
\end{figure}
\begin{figure}[thb]
\begin{center}
\pfig{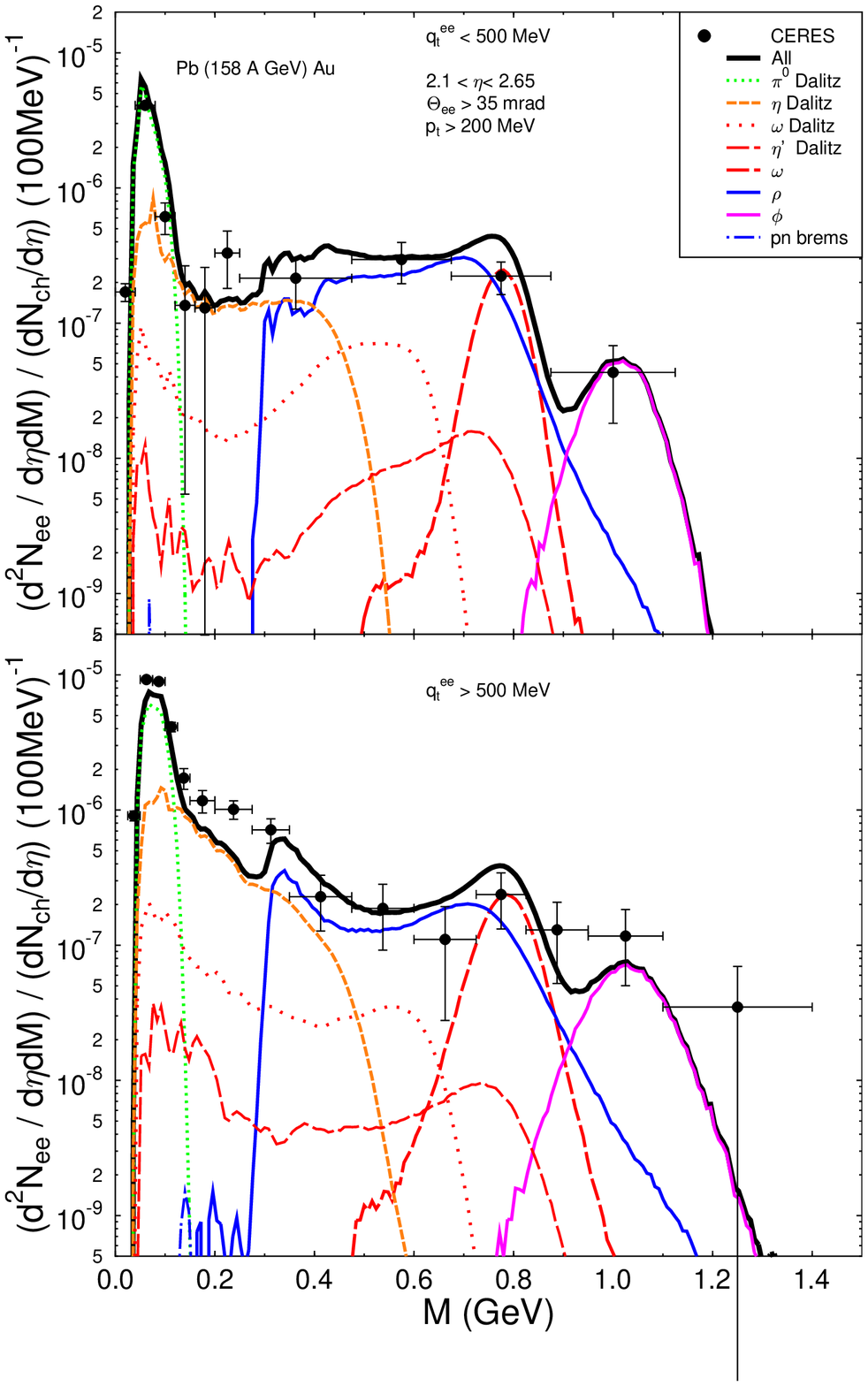}{11cm}
{Same as fig.~\ref{pbau} but with additional cuts on the
pair transverse momentum $q_t^{\epm}$ of the dilepton. The
low $p_t$ set is displayed in the upper part.}
\end{center}
\end{figure}

\end{document}